\documentclass[prl,twocolumn,preprintnumbers,amsmath,amssymb, nopacs,nokeys]{revtex4}

\usepackage{graphicx}% Include figure files
\usepackage{dcolumn}% Align table columns on decimal point
\usepackage{bm}% bold math
\usepackage{mathrsfs}
\usepackage{subfigure}
\usepackage{natbib}
\usepackage{amsmath}
\usepackage{amssymb}
\usepackage[english]{babel}
\graphicspath{{./figures/}}

\begin{document}
\title{Odd-frequency superconductivity in Sr$_2$RuO$_4$ measured by Kerr rotation}

\author{L. Komendov\'{a}}

\author{A. M. Black-Schaffer}
\affiliation{Department of Physics and Astronomy, Uppsala University, Box 530, SE-751 21 Uppsala, Sweden}

%\pacs{74.45.+c, 74.20.Mn, 74.20.Rp, 74.50.+r} %pacs numbers classification, to fill in before submitting
%\keywords{Multiband odd-frequency superconductivity; interband pairing; hybridization; p-wave; Kerr effect}
\date{\today}

\begin{abstract}
We establish the existence of bulk odd-frequency superconductivity in Sr$_2$RuO$_4$ and show that an intrinsic Kerr effect is a direct evidence of this state.  We use both general two- and three-orbital models, as well as a realistic tight-binding description of Sr$_2$RuO$_4$ to demonstrate that odd-frequency pairing arises due to finite hybridization between different orbitals in the normal state, and is further enhanced by finite inter-orbital pairing. 
\end{abstract}

\maketitle
% --------------------------------------- %
% INTRODUCTION:
% --------------------------------------- %
The layered perovskite strontium ruthenate Sr$_2$RuO$_4$ hosts an exotic superconducting state at low temperatures \cite{Maeno1994}, with experiments having established both spin-triplet  pairing \cite{Ishida1998, Duffy2000} and broken time-reversal symmetry \cite{Luke1998, Kapitulnik-Kerr2006}. The strongest candidate for the spatial symmetry is chiral $p$-wave order, with $d$-vector $\mathbf{d} = \Delta_0 \hat{\mathbf{z}} (k_x \pm i k_y)$ \cite{Rice1995, Mackenzie2003}, although the exact gap structure is still disputed \cite{Maeno2012, Kallin-review2009, Kallin-review2012, Kallin-review2016}. For example, spontaneous edge currents, associated with chiral $p$-wave pairing, have so far not been found \cite{Curran2014}. The situation is further complicated by the Fermi surface consisting of three bands originating from three hybridizing Ru $4d$ orbitals; electron-like $\gamma$- (from $xy$ orbital) and $\beta$-bands ($xz$, $yz$ orbitals) and hole-like $\alpha$-band ($xz$ and $yz$ orbitals).   

In this work we establish that the multi-orbital nature of Sr$_2$RuO$_4$ hosts bulk odd-frequency (odd-$\omega$) superconductivity.
Odd-$\omega$ superconductivity is a dynamical phenomenon where the fermionic nature of the Cooper pair is preserved due to an oddness in frequency (or equivalently time) \cite{Berezinskii, Balatsky1992}. It is well established to exist in superconductor-ferromagnet junctions \cite{Bergeret, reviewBergeret, Robinson2010, Bernardo2015, Triola2016} and also predicted for superconductor-normal metal junctions \cite{Tanaka2007a, Tanaka2007b, Tanaka2007c}. However, odd-$\omega$ superconductivity has remained elusive in bulk materials without external magnetic fields. Recently, an intriguing possibility of bulk odd-$\omega$ pairing was proposed for model two-band superconductors. Here an odd parity in the band index of the Cooper pair induces an odd-$\omega$ dependence \cite{AnnicaMB, Komendova2015}. We demonstrate odd-$\omega$ superconductivity in Sr$_2$RuO$_4$ arising from a similar odd parity in the orbital index.

Importantly, we also show that a finite Kerr rotation angle in clean Sr$_2$RuO$_4$ is direct evidence of odd-$\omega$ superconductivity. 
Detecting the Kerr effect in superconducting Sr$_2$RuO$_4$ was instrumental for establishing time-reversal symmetry breaking \cite{Kapitulnik-Kerr2006}. Recently, it has also been shown that inter-orbital processes are needed for a finite Kerr rotation \cite{Taylor2012, Wysokinski2012, Gradhand2013, Taylor2013}, unless invoking extrinsic impurity effects \cite{Goryo2008, Lutchyn2009}. We find odd-$\omega$ superconductivity and Kerr rotation emerging from the same finite hybridization between different orbitals and supplemented by possible finite inter-orbital pairing. 

% ------------------ %
% RESULTS:
% ------------------ %
% ------------------ %
% TWO ORBITALS
% ------------------ %
{\it Two-orbital superconductor.---}
To establish odd-$\omega$ superconductivity in Sr$_2$RuO$_4$ and its connection to the Kerr effect we start by studying a minimal two-orbital model. The general Bogoliubov-de Gennes Hamiltonian reads
$\sum_{\mathbf{k}} \Psi^\dagger_{\mathbf{k}}  {\hat H}_{\mathbf{k}} \Psi_{\mathbf{k}} $, where
\begin{equation}
\label{blockH}
{\hat H}_{\mathbf{k}} = \begin{pmatrix}
{\hat H}_0(\mathbf{k}) & \check{\Delta} (\mathbf{k}) \\
\check{\Delta}^{\dagger}(\mathbf{k}) & -{\hat H}_0(-\mathbf{k})  
\end{pmatrix}
\end{equation}
and $\Psi^\dagger_{\mathbf{k}} = (c^\dagger_{\mathbf{k} \uparrow 1} \, c^\dagger_{\mathbf{k} \uparrow 2} \,
c_{-\mathbf{k} \downarrow 1} \, c_{-\mathbf{k} \downarrow 2})$ for orbitals 1 and 2.
The normal and superconducting parts of the Hamiltonian are, respectively,
\begin{equation}
{\hat H}_0(\mathbf{k}) = \begin{pmatrix}
\xi_1 & \epsilon_{12} \\
\epsilon_{12} & \xi_2 
\end{pmatrix}, \,\,\,
\check{\Delta} (\mathbf{k}) = \begin{pmatrix}
\Delta_1 & \Delta_{12} \\
\Delta_{12} & \Delta_2
\end{pmatrix}.
\end{equation}
We here make only two assumptions: the normal state is even in momentum and spin-degenerate, and the superconducting state has only opposite spin pairing, i.e.,~either spin-singlet or spin-triplet pairing with $\mathbf{d} \parallel \hat{\mathbf{z}}$ in all orbitals. This clearly includes all chiral $p$-wave models used for Sr$_2$RuO$_4$.
Furthermore, to respect hermiticity, the inter-orbital hybridization $\epsilon_{ij} = \epsilon_{ji}$ and inter-orbital pairing $\Delta_{ij} = \Delta_{ji}$. We note here that by diagonalizing only ${\hat H}_0(\mathbf{k})$, we pass from the original orbital basis to, what we denote, the band basis. %However, such process will induce additional inter-band pairing terms dependent on the inter-orbital hybridization.

Using the Hamiltonian in Eq.~\eqref{blockH} we calculate the matrix Green's function $\check{G} = (i\omega-{\hat H})^{-1}$, where $\omega$ is the fermionic Matsubara frequency. The resulting matrix has the structure (arrow denoting the hole propagator):
\begin{equation}
\check{G} =
\begin{pmatrix}
G_{i j} & F_{i j} \\  
F_{i j}^\dagger & \overleftarrow{G}_{i j}     \\      
\end{pmatrix}.
\end{equation}
For odd-$\omega$ symmetry only the inter-orbital pairing amplitudes $F_{12}$ and $F_{21}$ are relevant, where the fixed parity states are the odd and even combinations
\begin{align}
\label{eq:Fodd}
 F_{12}-F_{21} = & i\omega [(\Delta_2- \Delta_1) \epsilon_{12} + \Delta_{12}(\xi_1 - \xi_2)]/D_2 \\
F_{12}+F_{21} = & [\epsilon_{12}(\Delta_2 \xi_1 + \Delta_1 \xi_2) + \Delta_{12}^*(\Delta_1 \Delta_2 -\Delta_{12}^2) \nonumber \\
-& \Delta_{12}(\omega^2 + \epsilon_{12}^2 + \xi_1 \xi_2)]/D_2, 
\end{align}
with $D_2$ a polynomial in $\omega^2$.
We see directly that the odd orbital combination, $F_{12}-F_{21}$, is also odd in frequency, while the even orbital combination is even in frequency. Both inter-orbital pairing terms retain the spatial and spin symmetries of the original pairing terms. There is thus a complete reciprocity in parity between orbital index and frequency, such that the Cooper pairs keep their required fermionic nature.
Moreover, the odd-$\omega$ component is finite only if there is finite inter-orbital pairing $\Delta_{12}$ or finite inter-orbital hybridization $\epsilon_{12}$. That either of these generate odd-$\omega$ pairing is not fully unexpected, since inter-band hybridization has been found to induce inter-band pairing \cite{Komendova2015}.
In addition, an asymmetry in intra-orbital properties is needed for odd-$\omega$ superconductivity to appear, such that either $\xi_1 \neq \xi_2$ (for finite $\Delta_{12}$) or $\Delta_1 \neq \Delta_2$ (for finite $\epsilon_{12}$). However, such asymmetry is almost always present in a real material. 

The expression in Eq.~\eqref{eq:Fodd} for odd-$\omega$ superconductivity can directly be compared to the optical Hall conductivity $\sigma_H(\omega)$. For the same two-orbital model Taylor and Kallin \cite{Taylor2012, Taylor2013} recently found both the real and imaginary parts of $\sigma_H(\omega)$ proportional to the factor $[\epsilon_{12} \mathrm{Im}(\Delta^*_{1}\Delta_{2}) + \xi_1 \mathrm{Im}(\Delta^*_{2}\Delta_{12}) - \xi_2 \mathrm{Im}(\Delta^*_{1}\Delta_{12})]$. With the Kerr rotation angle proportional to $\textrm{Im}[\sigma_H(\omega)]$, it is clear that time-reversal symmetry needs to be broken for a finite Kerr effect. But most importantly here, a finite Kerr rotation also crucially requires either inter-orbital hybridization or inter-orbital pairing, which are exactly the same basic requirements as for odd-$\omega$ superconductivity. Moreover, the needed intra-orbital asymmetry for finite odd-$\omega$ pairing is also required for the Kerr effect. 
For finite inter-orbital pairing $\Delta_{12}$, non-degenerate orbital dispersions, which is usually always the case, are needed for both finite Kerr rotation and odd-$\omega$ pairing.
In the case of finite inter-orbital hybridization $\epsilon_{12}$, a more detailed analysis has found that an asymmetry between the intra-orbital pairings $\Delta_{1,2}$ is absolutely crucial for a finite Kerr angle \cite{Mineev, Gradhand2013}, which is exactly as for finite odd-$\omega$ pairing. We therefore conclude that the presence of an intrinsic Kerr effect is direct evidence of bulk odd-$\omega$ superconductivity. Conversely, odd-$\omega$ superconductivity in a time-reversal breaking superconductor can be directly probed by the Kerr effect.
   
% ------------------ %
% THREE ORBITALS
% ------------------ %
{\it Three-orbital superconductor.---} 
To fully address the physics of Sr$_2$RuO$_4$ we need to go beyond two-orbitals and include all three orbitals present at the Fermi level. Following the same notation and assumptions as before, a general three-orbital superconductor is written as
\begin{equation}
{\hat H}_0(\mathbf{k}) = \begin{pmatrix}
\xi_1 & \epsilon_{12} & \epsilon_{13} \\
\epsilon_{12} & \xi_2 & \epsilon_{23} \\
\epsilon_{13} &  \epsilon_{23} & \xi_3 \\
\end{pmatrix} , \,\,\,
\check{\Delta} (\mathbf{k}) = \begin{pmatrix}
\Delta_1 & \Delta_{12} & \Delta_{13} \\
\Delta_{12} & \Delta_2 & \Delta_{23} \\
\Delta_{13} & \Delta_{23} & \Delta_3
\end{pmatrix}. \nonumber
\end{equation}
We again extract the\ inter-orbital pairing amplitudes from the Green's function $\check{G} = (i\omega-{\hat H})^{-1}$. However, with three orbitals we first need to generalize the concept of parity in orbital index. For this we use the completely antisymmetrized $F_{AS} = \sum_{i,j,k = 1,\ldots,N} \epsilon_{ijk}F_{ij}$, which for $N = 3$ orbitals results in $F_{AS} = F_{12} - F_{21} + F_{23} - F_{32} + F_{31} - F_{13}$, and the symmetric inter-orbital pairing $F_{S} = \sum_{i \neq j = 1,\ldots, N} F_{ij}$. $F_S$ and $F_{AS}$ naturally keep the spatial and spin symmetries of their inter-orbital parents. We have confirmed that for a general three-orbital system $F_{AS} (-\omega)= -F_{AS}(\omega)$ and $F_{S} (-\omega)= F_{S}(\omega)$, i.e.,~with these definitions there is a full reciprocity between parity in orbital index and frequency, just as in the two-orbital case. Having this established, we from now on refer to $F_{S}$ as $F_{\mathrm{even}}$ and  $F_{AS}$ as $F_{\mathrm{odd}}$, with the subscript reflecting both parity in orbital index and frequency.

We can analytically draw several conclusions for a general three-orbital superconductor by using a few simplifying assumptions. First we consider the case of only intra-orbital pairing, i.e.~$\check{\Delta} (k) = \mathrm{diag} (\Delta_{1},\Delta_{2},\Delta_{3})$, while keeping a generic inter-orbital hybridization by setting $\epsilon_{ij} = \Gamma$ for all $i \neq j = 1,2,3$. We then find 
\begin{align}
\label{eq:Fodd3a}
F_{\mathrm{odd}} & = 2 \Gamma i \omega [\Delta_1 (\epsilon_2 - \epsilon_3) (\epsilon_2 + \epsilon_3 + \Gamma)  + |\Delta_1|^2(\Delta_3 - \Delta_2) \nonumber\\
& + \mathrm{two} \, \mathrm{cyclic} \,\mathrm{permutations}]/D_3,
\end{align}
where $D_3$ is a polynomial in $\omega^2$. The odd-$\omega$ component is thus directly proportional to the inter-orbital hybridization $\Gamma$, but it also requires asymmetry in the intra-orbital parameters, $\epsilon_2 \neq \epsilon_3$ or $\Delta_3 \neq \Delta_2$ etc.. This demonstrates that odd-$\omega$ pairing does not require any intrinsic inter-orbital pairing, but only finite inter-orbital hybridization, making it ubiquitous in three-orbital superconductors.

Alternatively, we can assume strong inter-orbital processes only between orbitals 2 and 3, keeping both $\epsilon_{23}$ and $\Delta_{23}$ finite, while all other inter-orbital terms are zero. This models the situation in Sr$_2$RuO$_4$, since the $xy$ and $yz$ orbitals hybridize strongly, eventually forming the $\alpha$- and $\beta$-bands, while the $xy$ orbital forms the $\gamma$- band without other orbitals contributing significantly. We then arrive at 
\begin{align}
\label{eq:Fodd3b}
F_{\mathrm{odd}} = 2 i \omega [\Delta_{23}(\epsilon_3-\epsilon_2)+\epsilon_{23}(\Delta_3 - \Delta_2)]/D'_3,
\end{align}
 with $D'_3$ a polynomial in $\omega^2$. This result is analogous to that of a two-orbital superconductor, which is  not surprising since orbital 1 is disconnected from the other two orbitals. Very interestingly, odd-$\omega$ pairing is in this case only destroyed if there is no inter-orbital pairing present, i.e.,~all $\Delta_{ij} = 0$, {\it and} the intra-orbital pairing in orbitals 2 and 3 is equal, $\Delta_2 = \Delta_3$. Exactly the same stringent conditions on the pairing state have recently been reported to be needed in order to destroy the intrinsic Kerr effect in three-orbital models of Sr$_2$RuO$_4$ (here assuming that $\epsilon_{12}$ is always finite) \cite{Mineev, Gradhand2013}. Moreover, with these conditions all fulfilled, a finite Kerr effect was only found to be restored when invoking a finite hybridization with orbital 1 ($xy$ orbital) \cite{Gradhand2013}. Assuming finite $\epsilon_{1j}$, here taken as $\epsilon_{12}=\epsilon_{13}$ for simplicity, we then also find the odd-$\omega$ component returning:
%\begin{align}
$F_{\mathrm{odd}}  \sim 2 i \omega \epsilon_{12}(\Delta_1 - \Delta_2) (\epsilon_2 - \epsilon_3) (\epsilon_{12} + \epsilon_2 + \epsilon_3)$.
%\end{align}
Since all orbitals in Sr$_2$RuO$_4$ have different dispersions $\epsilon_i$, and the gap relation $\Delta_{1} \neq \Delta_{2,3}$ is widely assumed, this shows that odd-$\omega$ pairing is necessarily present in Sr$_2$RuO$_4$ and also probed by the intrinsic Kerr effect. This is true even without having to invoke any finite inter-orbital pairing.

% ------------------ %
% Sr$_2$RuO$_4$
% ------------------ %
% FIGURE 1:
	\begin{figure*}[t]
		\includegraphics[width=0.9\textwidth]{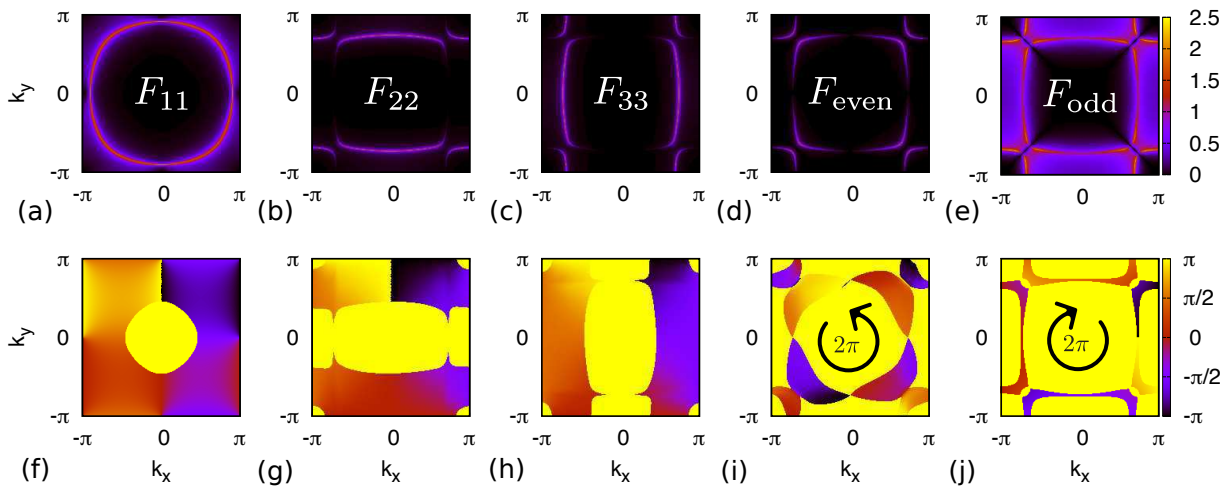}
		\caption{Anomalous Green's functions for $\eta_{1} = 5$, $\eta_{2}=\eta_{3}=2.5$, and $\eta_{23} = 0.5$ with amplitudes (top) and complex phases (bottom), divided into (a) $F_{11}$ ($d_{xy}$), (b) $F_{22}$ ($d_{xz}$), (c) $F_{33}$ ($d_{yz}$), and inter-orbital (d) $F_{\mathrm{even}}$, and (e) $F_{\mathrm{odd}}$. The amplitude of $F_{\mathrm{odd}}$ is multiplied by 100.  
			\label{fig1}}
	\end{figure*}
{\it Odd-$\omega$ superconductivity in Sr$_2$RuO$_4$.---} We finally turn to a full numerical calculation of the odd-$\omega$ pairing in Sr$_2$RuO$_4$ using a realistic tight-binding model \cite{Gradhand2013} based on de Haas-van Alphen Fermi surface data \cite{Bergemann2000}. For simplicity we ignore the small dispersion in $z$, which weakly couples the $xy$ orbital with the other two orbitals, and focus on $k_z =0$. Labeling the three ruthenium $d$ orbitals $(xy, xz, yz)$ by (1, 2, 3) we have
\begin{align}
\xi_1 (\mathbf{k})& =  \epsilon_1 + 2t(\cos k_x+\cos k_y) + 4 t^\prime \cos k_x \cos k_y \nonumber \\ 
\xi_2 (\mathbf{k}) & =  \epsilon + 2(t_2 \cos k_x + t_3 \cos k_y) + \epsilon^{\perp} (\mathbf{k})\nonumber \\
\xi_3 (\mathbf{k}) & =  \epsilon + 2(t_3 \cos k_x + t_2 \cos k_y) + \epsilon^{\perp} (\mathbf{k})\nonumber \\
\epsilon^{\perp} (\mathbf{k}) & = 8 t^{\perp} \cos (k_x/2) \cos (k_y/2) \nonumber \\
\epsilon_{23} (\mathbf{k}) & = 4 t_{23} \sin k_x \sin k_y + 8 t_{23}^{\perp} \sin (k_x/2) \sin (k_y/2) \nonumber \\ 
\epsilon_{12} (\mathbf{k}) & = \epsilon_{13} (\mathbf{k}) =0 \nonumber,
\end{align}
with the parameters $(\epsilon_1, \epsilon) = (-131.8, -132.22)$, $(t, t^\prime, t_2, t_3)= (-81.62, -36.73, -109.37, -6.56)$, and $(t^\perp, t_{23}, t_{23}^\perp) = (0.262, -8.75, -1.05)$.
The superconducting state is further described by
\begin{align}
\Delta_1 (\mathbf{k}) =& \eta_1 (\sin k_x + i \sin k_y) \nonumber \\
\Delta_2 (\mathbf{k}) =& \eta_2 \sin (k_x/2) \cos (k_y/2) + i\eta_3 \cos (k_x/2) \sin (k_y/2) \nonumber \\
\Delta_3 (\mathbf{k}) =& \eta_3 \sin (k_x/2) \cos (k_y/2) + i\eta_2 \cos (k_x/2) \sin (k_y/2) \nonumber\\
\Delta_{23} (\mathbf{k}) =& \eta_{23} [\sin (k_x/2) \cos (k_y/2) - i \cos (k_x/2) \sin (k_y/2) ] \nonumber\\
\Delta_{12} (\mathbf{k}) =& \Delta_{13} (\mathbf{k}) =0. \nonumber
\end{align}
Thus, the only inter-orbital terms are the hybridization $\epsilon_{23}$ and pairing $\Delta_{23}$ between the $(xz,yz)$ orbitals, which give rise to the $\alpha$- and $\beta$-bands. From our general results above we know that either of these terms gives odd-$\omega$ pairing directly proportional to the corresponding inter-orbital term.
Our parametrization also means that the $xy$-orbital, forming the $\gamma$-band, is not interacting with the other two orbitals. We thus ignore the additionally small, but existing, odd-$\omega$ pairing contribution arising from finite contributions of the $xz,yz$ orbitals into the $\gamma$-band, which is present away from $k_z =0$.
Also note that the interactions in the $x$- and $y$-directions for the ($xz, yz$) orbitals are not related by symmetry, and thus $\eta_2 \neq \eta_3$ is generally allowed. Such a difference automatically leads to an asymmetry between the intra-orbital pairing in orbitals 2 and 3, since $\Delta_3 - \Delta_2 \propto \eta_3 - \eta_2$. Thus, asymmetry between the $x$ and $y$ direction for each orbital translates directly into an intra-orbital pairing asymmetry. 
Our model also ignores spin-orbit coupling, which has been shown to be rather strong in Sr$_2$RuO$_4$ (130$\pm$30 meV) \cite{Veenstra2014}. However, this will primarily only influence the spin structure of the superconducting state, while the odd-$\omega$ state arises from the combined symmetry under orbital and time interchange. As such, the odd-$\omega$ pairing is independent on the spin and spatial symmetries of the superconducting state. Thus we do not expect finite spin-orbit coupling to qualitatively change any of our results.

Since superconductivity in Sr$_2$RuO$_4$ has been found to be either dominant on the $\gamma$-band \cite{Deguchi2004}, comparable \cite{Scaffidi2014}, or dominant on the $\alpha$- and $\beta$-bands \cite{Raghu2010}, we opt to not solve self-consistently for any of the pairing gaps. Rather, we tune their relative sizes in each of the three orbitals, also including possible inter-orbital pairing $\Delta_{23}$, which then spans the cases discussed in literature. We extract a $k$-space structure by summing $F_{\mathrm{odd}}$ over all positive Matsubara frequencies \footnote{In practice, we find it sufficient use 200 as the upper bound using a step size of 1 for the numerical summations to converge. Also, the inclusion of $\omega = 0$ does not play a role for $F_{\mathrm{odd}}$, since it per definition zero.}. For all results we set 1 meV as the base unit, but for better resolution in the plots we use values of the order parameters $\eta$ about 20 times higher than expected. However, we have checked that the physics remains unchanged using more realistic values. 

First we discuss the results for one representative set of order parameters, including finite inter-orbital pairing. In Fig.~\ref{fig1} we show the pairing amplitudes $F_{11}$, $F_{22}$ and $F_{33}$ in panels (a)-(c), and their corresponding complex phases in (f)-(h). These are the intra-orbital pairings in each Ru $4d$ orbital: $xy$ ($\gamma$-band), $xz$ and $yz$. We also display the even and odd inter-orbital pairing amplitudes $F_{\mathrm{even}}$ (d,i) and $F_{\mathrm{odd}}$ (e,j). As expected, the inter-orbital pairing amplitudes are located just on the $\alpha$- and $\beta$-bands. The odd-$\omega$ inter-orbital pairing acquires nodal lines close to the Brillouin zone corners, where the phase of $F_{\mathrm{odd}}$ changes abruptly. However, despite four nodal lines, the pairing remains odd in $\mathbf{k}$. The phase of $F_{\mathrm{even}}$ also shows an interesting rotation by $\pi$ between the $\alpha$- and $\beta$-bands. For both $F_{\mathrm{even}}$ and $F_{\mathrm{odd}}$ the overall phase winds by $2 \pi$  around the $\Gamma$ point, as expected for a chiral $p$-wave state. Note that the phase however grows opposite to this direction within each individual lobe, while still keeping an overall $2\pi$ rotation.    

\begin{figure}[htb]
	\includegraphics[width=0.4\textwidth]{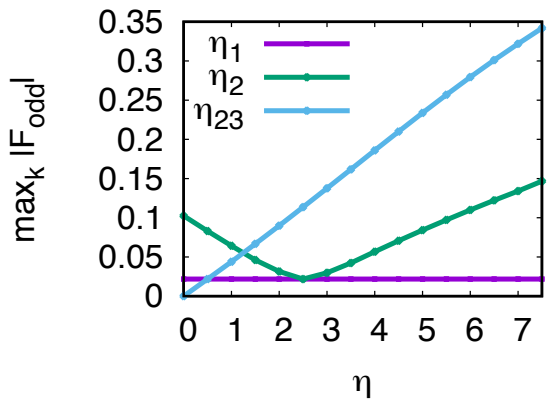}
	\caption{Maximum amplitude of $F_{\mathrm{odd}}$ with changing order parameters $\eta_{1}$,  $\eta_{2}$, and $\eta_{23}$. Non-changing order parameters are fixed to $\eta_{1} = 5$, $\eta_{2}=\eta_{3}=2.5$, and $\eta_{23} = 0.5$. 
		\label{fig2}}
\end{figure}

Next we explore how the maximum odd-$\omega$ superconducting amplitude is influenced by changing the different order parameters. From Fig.~\ref{fig2} we find that the odd-$\omega$ pairing is directly proportional to both the inter-orbital term $\eta_{23}$ and the difference $|\eta_{2}-\eta_{3}|$, while changing $\eta_1$ has no influence. 
This is in agreement with the results presented in Eq.~\ref{eq:Fodd3b}, since inter-orbital processes are only present between orbitals 2 and 3, while orbital 1 is disconnected. Thus odd-$\omega$ pairing in Sr$_2$RuO$_4$ is enhanced either by increasing the inter-orbital pairing $\eta_{23}$ or increasing the asymmetry between the intra-orbital pairing in orbitals 2 and 3, with the latter enhancing the influence of the (always finite) inter-orbital hybridization $\epsilon_{23}$.
%
% FIGURE 2:

\begin{figure}
	\includegraphics[width=0.8\linewidth]{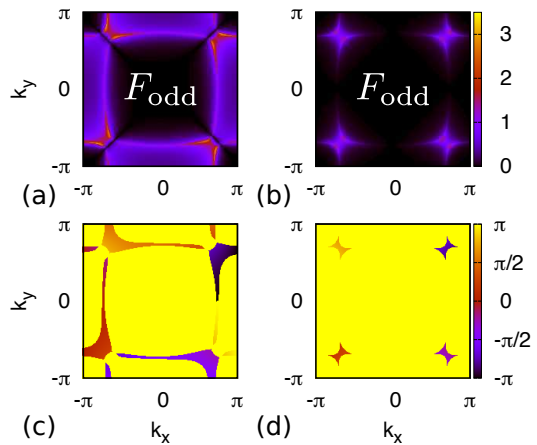}
	\caption{Anomalous Green's function $F_{\mathrm{odd}}$ with amplitudes (top) and complex phases (bottom) for finite intra-orbital pairing asymmetry using $\eta_2=2$ and $\eta_3=3$ with (a,c) inter-orbital pairing $\eta_{23} = 0.5 $ and (b,d) $\eta_{23} = 0$. Also $\eta_1 = 5$.   
		\label{fig3}}
\end{figure}

Having established how the maximum odd-$\omega$ amplitude depends on the inter-orbital processes in Sr$_2$RuO$_4$, we finally focus on how its $k$-space structure evolves when changing these parameters. In Fig.~\ref{fig3} we introduce a finite asymmetry in the intra-orbital pairing in the $xz$- and $yz$-orbitals, i.e.~$\eta_2 - \eta_3 \neq 0$. Keeping the same inter-orbital pairing as in Fig.~\ref{fig1}, we see in Fig.~\ref{fig3}(a,c) how the mirror symmetry with respect to the $|k_x|=|k_y|$ diagonals is lifted when introducing the additional odd-$\omega$ pairing coming from the inter-orbital hybridization. Changing the sign of the $(\eta_2-\eta_3)$ difference mirror-reflects the figure. With even larger disparity between the intra-orbital pairing parameters, the nodal lines are even visibly shifted away from the diagonals such that they do not cross at the $\Gamma$-point. The overall $2\pi$ phase winding of the chiral $p$-wave symmetry is however always present.
%
% FIGURE 3:

We then study the same situation ($\eta_2 - \eta_3 \neq 0$), but now turn off the inter-orbital pairing $\eta_{23}$. In this case, displayed in Fig.~\ref{fig3}(b,d), the odd-$\omega$ pairing is strongly localized to the intersecting points between orbitals 2 and 3 and there are no visible nodal lines. Changing the sign of $(\eta_{2}-\eta_{3})$ rotates the phase of $F_{\mathrm{odd}}$ by $\pi$. Since the odd-$\omega$ pairing in this case is entirely driven by the inter-orbital hybridization $\epsilon_{23}$, it is not surprising that the odd-$\omega$ is also highly peaked in the regions with largest hybridization. If we also set $\eta_{2}=\eta_{3}$ we lose the odd-$\omega$ signal altogether (and the intrinsic Kerr effect, see above). 
The overall structure of the other anomalous Green's functions, $F_{11}$, $F_{22}$, $F_{33}$, and $F_{\mathrm{even}}$, remains largely unaffected by changing $(\eta_2-\eta_3)$ or $\eta_{23}$. 

% ------------------ %
% SUMMARY:
% ------------------ %
In summary, we have established that bulk odd-$\omega$ superconductivity is ubiquitous in multi-orbital superconductors and in particular in Sr$_2$RuO$_4$. Odd-$\omega$ pairing in Sr$_2$RuO$_4$ arises due finite hybridization between the different ruthenium orbitals in the normal state, and is further enhanced by finite inter-orbital pairing. Using both effective two- and three-orbital models we have also demonstrated that an intrinsic Kerr effect is direct evidence for bulk odd-$\omega$ superconductivity. We find it likely that similar connections can be found in other time-reversal symmetry breaking systems, such as UPt$_3$ \cite{Schemm2014} or Bi/Ni bilayers \cite{Xinxin2016}.

\begin{acknowledgments} 
We thank A.~Balatsky and A.~Bouhon for helpful discussions. This work was supported by the Wenner-Gren Foundations, the Swedish Research Council (Vetenskapsr\aa det), the Swedish Foundation for Strategic Research (SSF), the G\"oran Gustafsson Foundation, and the Wallenberg Academy Fellows program.
\end{acknowledgments}

% BIBLIOGRAPHY:

\end{document}